\documentclass[aps,twocolumn,prl,amsmath,showpacs,superscriptaddress,floatfix]{revtex4}
\newcommand{\myfigscale}{0.57}

\usepackage{graphicx}
\usepackage{dcolumn}% Align table columns on decimal point
\usepackage{bm}% bold math

\newcommand{\jop}{j}
\newcommand{\linconst}{\gamma}

\begin{document}

\title{Lateral transport of thermal capillary waves}

\author{Thomas H. R. Smith}
\affiliation{H.H. Wills Physics Laboratory, University of Bristol,
  Tyndall Avenue, Bristol BS8 1TL, United Kingdom}
\author{Oleg Vasilyev}
\affiliation{Max-Planck-Institut f{\"u}r Metallforschung,
  Heisenbergstra\ss e~3, D-70569 Stuttgart, Germany}
\affiliation{Institut f\"ur Theoretische und Angewandte Physik,
Universit\"at Stuttgart, D-70569 Stuttgart, Germany}
\author{Anna Macio\l ek}
\affiliation{Max-Planck-Institut f{\"u}r Metallforschung,
  Heisenbergstra\ss e~3, D-70569 Stuttgart, Germany}
\affiliation{Institut f\"ur Theoretische und Angewandte Physik,
  Universit\"at Stuttgart, D-70569 Stuttgart, Germany}
\affiliation{Institute of Physical Chemistry, Polish Academy of Sciences,
  Department III, Kasprzaka 44/52, PL-01-224 Warsaw, Poland}
\author{Matthias Schmidt}
\affiliation{H.H. Wills Physics Laboratory, University of Bristol,
  Tyndall Avenue, Bristol BS8 1TL, United Kingdom}
\affiliation{Theoretische Physik II, Universit\"at Bayreuth, 
  Universit\"atsstra\ss e 30,  D-95440 Bayreuth, Germany}
\date{19 April 2009}

\begin{abstract}
   We demonstrate that collective motion of interfacial fluctuations
   can occur at the interface between two coexisting thermodynamic
   phases. Based on computer simulation results for driven diffusive
   Ising and Blume-Capel models, we conjecture that the thermal
   capillary waves at a planar interface travel along the interface if
   the lateral order parameter current $\jop(y)$ is an odd function of
   the distance $y$ from the interface and hence possesses opposite
   directions in the two phases. Such motion does not occur if
   $\jop(y)$ is an even function of $y$. A discrete Gaussian interface
   model with effective dynamics exhibits similiar transport phenomena
   but with a simpler dispersion relation. These findings open up
   avenues for controlled interfacial transport on the nanoscale.
\end{abstract}

\pacs{05.40.-a, 05.50.+q, 68.05.Cf, 68.35.Rh}

\maketitle

%% 05.40.-a Fluctuation phenomena, random processes, noise, and Brownian motion
%% 05.50.+q Lattice theory and statistics (Ising, Potts, etc.) 
%%   (see also 64.60.Cn Order-disorder transformations, and 
%%             75.10.Hk Classical spin models)
%% 68.05.Cf Liquid-liquid interface structure: measurements and simulations
%% 68.35.Rh Phase transitions and critical phenomena

Understanding the motion of interfaces is important in areas such as
multiphase flow, dendritic and crystal growth, microchip fabrication,
combustion, blood flow, and cell dynamics. Pronounced changes to the
microscopic interfacial structure can occur as a result of the motion,
e.g., the development of kinetically enhanced self-affine roughness
\cite{HZandRikvold} of the driven interface between a growing material
and its environment. In striking contrast, driving the interface along
the plane of its average position can lead to interfacial smoothening,
as has been observed experimentally for a colloidal gas-liquid
interface under shear flow \cite{derks}, and in computer simulations
of interfaces in driven lattice gas models~\cite{zia,SVAM}. Coherent
lateral transport of large-scale interfacial structures is well-known
in macroscopic systems that are far from equilibrium, such as
migrating sand dunes and ripples, or ocean waves. Finding thermal
analogues can be of relevance in micro- and nanofluidic devices,
e.g.\ due to the possibility of transport of nanoparticles at a
liquid-liquid interface by controlled motion of thermal capillary
waves.

Consider two distinct equilibrium phases at thermodynamic phase
coexistence, organised into two different regions of space that are
separated by a planar interface.  The local order parameter $\Phi$
varies along the axis $y$ perpendicular to the interface plane and for
large enough systems reaches its respective bulk value far from the
interface. The (scalar) order parameter profile $\Phi(y)$ can
represent the density near a gas-liquid interface, the relative
concentration at the interface separating two coexisting liquid
phases, or the magnetization near an Ising domain wall. On a
coarse-grained scale, the interface can be characterized by its local
departure (height) $h(x,t)$ from a reference plane $y=h(x,t)=0$, where
$x$ indicates the coordinate(s) parallel to the interface and $t$ is
time. At temperatures $T$ above the roughening transition, the
interface exhibits large spatial fluctuations. Phenomenological
capillary wave theory \cite{cap_wave}, as well as rigorous results
\cite{DBA}, indicate that in the absence of external fields that
couple to the order parameter, the length scale of such fluctuations
diverges with system size, i.e., the interface thickness becomes
infinite in the thermodynamic limit in spatial dimensions $d\leq 3$.

In this Letter we consider external driving that creates a steady
state with non-vanishing current of the order parameter, $\jop(y)$,
parallel to a planar interface. We investigate the effects on the
dynamics of the capillary waves via the two-point correlation function
$C(x,t)=\langle h(0,0)h(x,t)\rangle$, where the angles denote an
average in the steady state.  Based on computer simulations results
for $C(x,t)$ for various simple microscopic models, we conjecture that
the {\em lateral flux of the order parameter at a planar interface
  induces lateral motion of the thermal capillary waves, provided
  $\jop(y)$ is an odd function of the distance $y$ from the
  interface.}
%%MS mention that no movement for odd j
The spatial symmetry at equal times, $C(x,0)=C(-x,0)$, is then broken
for times $t>0$, such that $C(x,t)\neq C(-x,t)$.  For general forms of
$\jop(y)$ it is the odd component, $[\jop(y)-\jop(-y)]/2$, that
induces the motion. No motion occurs when $\jop(y)$ is even. We shed
light on the transport mechanism by constructing a corresponding
(Gaussian) effective interface model that displays similar transport
phenomena.

As a microscopic approach we use kinetic lattice models with dynamics
that conserve the order parameter locally. The Ising and the
Blume-Capel \cite{blume-capel} models of binary mixtures possess the
Hamiltonian ${\cal H}= -J\sum_{\langle i,l \rangle}\sigma_i\sigma_l$,
where $\langle i,l \rangle$ denotes nearest-neighbor pairs and the
spin-spin coupling constant $J>0$. Contributions to $\cal H$ from
static external fields are omitted, as we will work in an ensemble of
fixed numbers of spins $\sigma_i$ of each type. For the Ising model
$\sigma_i=\pm 1$, while for the Blume-Capel model $\sigma_i=-1,0,+1$,
which in lattice fluid language corresponds to occupancy of site $i$
by particles of species ``$-1$'', vacancy, or species ``$+1$'' of the
mixture, respectively.  We initially consider two-dimensional square
lattices of dimensions $L_x \times L_y$ with coordinates $x$
(horizontal) and $y$ (vertical). The interface with mean orientation
in the $x$-direction is established and localized by assuming boundary
conditions $\sigma_i= +1$ at the upper ($y=(L_y+1)/2$) and
$\sigma_i=-1$ at the lower ($y=-(L_y+1)/2$) edges of the lattice.
Periodic boundary conditions are applied in the $x$ direction.  As a
consequence, $\Phi(y)=\langle \sum_x\sigma_i\rangle/L_x$, $i=(x,y)$
crosses over from $\Phi<0$ for $y<0$ to $\Phi>0$ for $y>0$.  We use
Kawasaki spin exchange dynamics \cite{kawasaki}, where the elementary
move consists of swapping the values of the spin variables $\sigma_i$
and $\sigma_l$ of two nearest-neighbor sites $i$ and $l$.  The system
is driven by a force field $F(y)$ that acts in the $x$-direction
(parallel to the interface) and varies with distance $y$ from the
interface. The acceptance rates for the trial moves are assumed to be
of modified Metropolis type, $\min\{1,\exp(-(\Delta H+\Delta
W)/(k_BT))\}$, where $\Delta H$ is the change in internal energy,
$\Delta W$ is the work due to the driving field, and $k_B$ is the
Boltzmann constant.  Hence the work performed by the field is
dissipated into a heat bath, which is kept at $T={\rm const}$.  We
have carried out extensive Monte Carlo (MC) simulations using
multi-spin \cite{gemmert05} coding techniques extended for Kawasaki
dynamics to include drive.  Results are presented for $L_x=200$ and
$L_y=20$ at fixed total magnetization $\sum_i\sigma_i=0$ for
$T/T_c=0.75$ (where $T_c=2.2692J/k_B$) and run lengths of the order of
$N_{\rm MC}=10^8$ MC steps ($L_x\times L_y$ trial moves).

We first discuss the Ising lattice gas where the drive creates a work
term for an exchange in the $x$-direction, $\Delta W= -J
F(y)(\sigma_i-\sigma_l)/2$, where $i=(x,y)$ and $l=(x+1,y)$; exchanges
in the $y$-direction occur with normal equilibrium rates, $\Delta
W=0$.  In the case of odd symmetry upon spatial reflection,
$F(y)=-F(-y)$, the field acts in opposing directions in both halves of
the system, e.g.\ with linear variation across the slit, $F(y) \equiv
\linconst y$, where $\linconst$ is the (scaled) field difference
between rows. Simulation results indicate that the order parameter
current profile $\jop(y)=j_+(y)-j_-(y)$, where $j_\sigma(y)$ is the
particle current profile of species $\sigma=\pm 1$, possesses the same
direction and the same symmetry as the driving field. For odd driving
fields, we find $\jop(y)=-\jop(-y)$. We define the instantaneous
interface position via a coarse-graining method based on the (scaled)
column magnetization, $h(x,t)=-(2m_{\rm b})^{-1}\sum_y \sigma_i$,
where site $i=(x,y)$ possesses the value $\sigma_i$ at time $t$, and
$m_{\rm b}$ is the spontaneous equilibrium magnetization in bulk.

\begin{figure}
  \includegraphics[scale=\myfigscale]{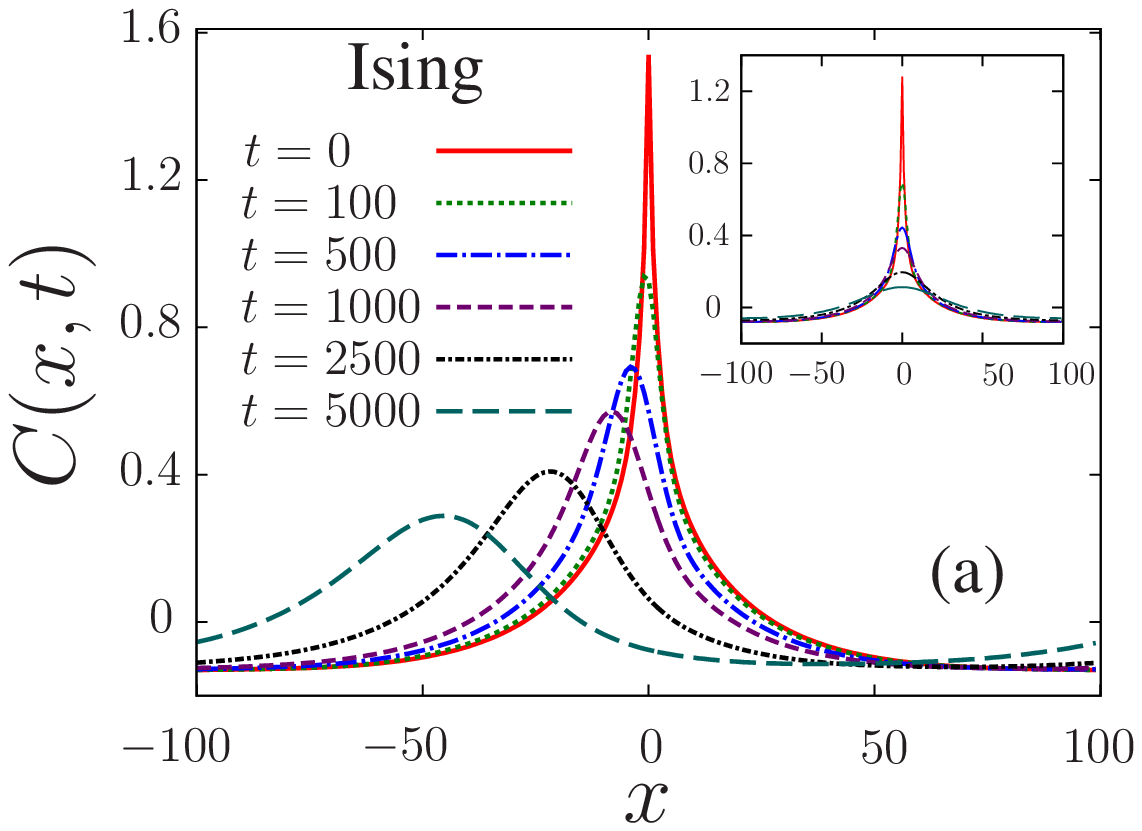}
  \includegraphics[scale=\myfigscale]{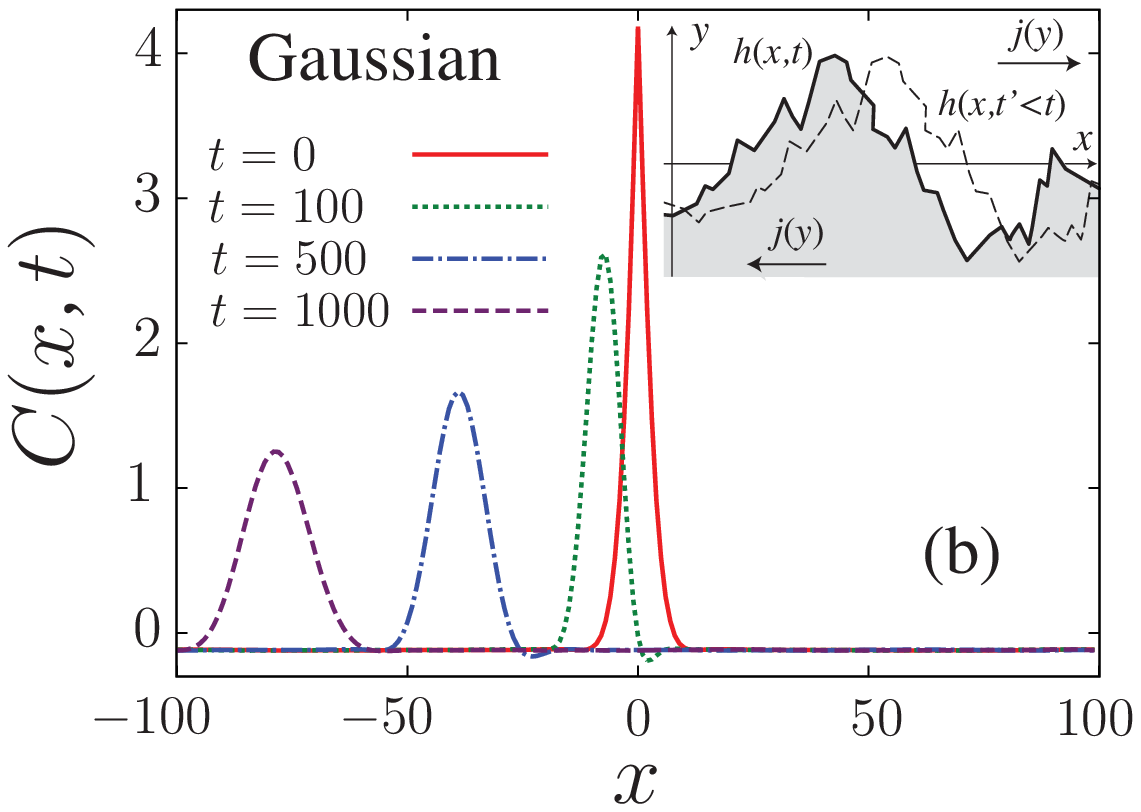}
  \caption{(color online) a) Height-height correlation function
    $C(x,t)$ of the Ising model as a function of the distance $x$ at
    constant time $t$ (as indicated, in units of MC steps) for
    temperature $T/T_c=0.75$, and system size $L_x=200$, and
    $L_y=20$. The system is subject to linearly varying drive with
    strength $\linconst=1$.  Inset: $C(x,t)$ for the case of V-shaped
    drive for the same parameters. b) Same as a), but for the discrete
    Gaussian interface model subject to linear $h$-dependent
    drive. Inset: Illustration of the motion of the fluctuating
    interface.}
  \label{fig:cxt}
\end{figure}
\begin{figure}
\vspace{-5mm}
  \includegraphics[scale=\myfigscale]{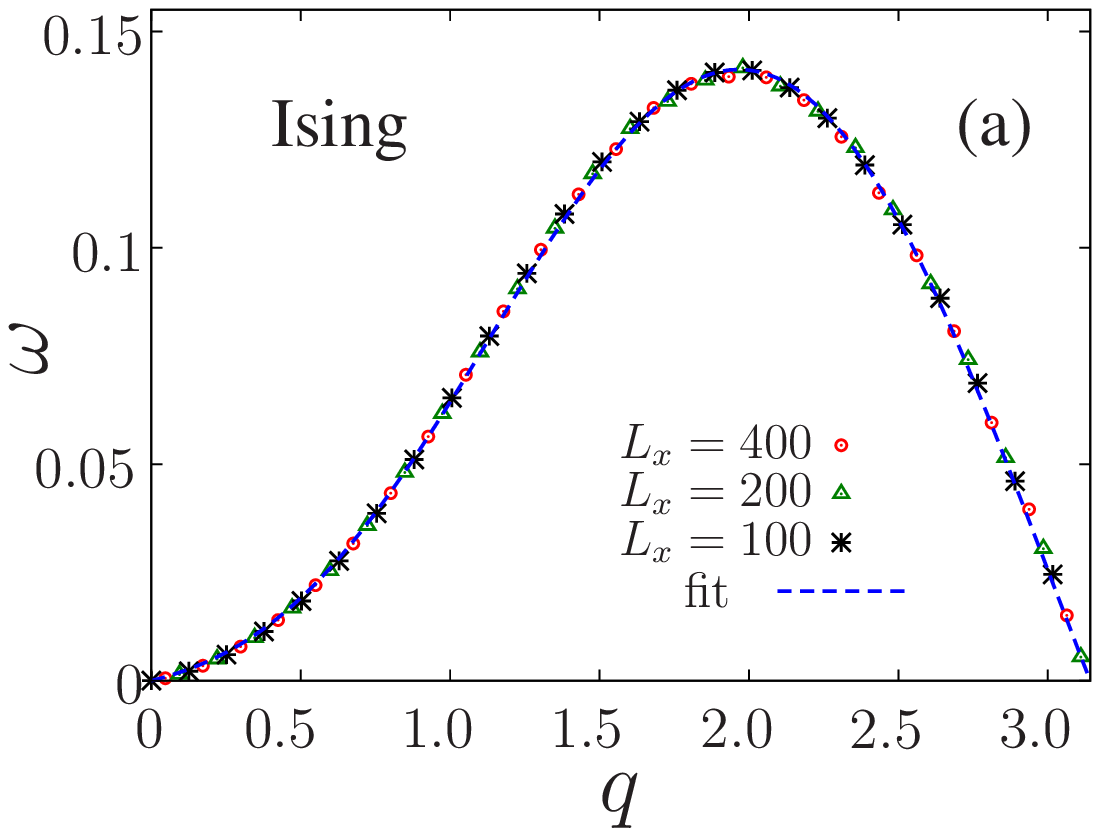}
  \includegraphics[scale=\myfigscale]{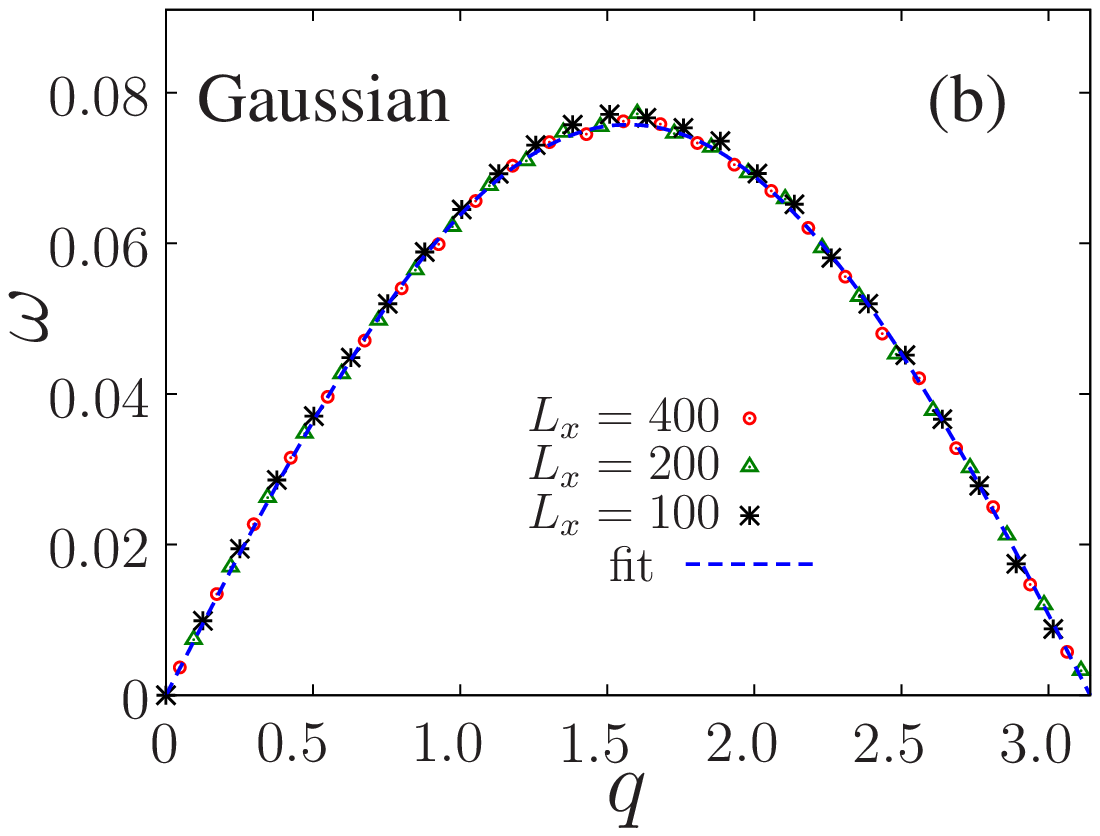}
  \caption{(color online) Dispersion relation $\omega(q)$ for the
    lateral propagation of capillary waves in the two-dimensional
    Ising model driven by a linear field for $\linconst=1$ (a), and in
    the one-dimensional kinetic discrete Gaussian model with
    height-dependent driving (b). The spectral variable is $q=2\pi
    k/L_x$, where $k$ is the wave number. Parameters are the same as
    in Fig.\ \ref{fig:cxt}. Data for different system sizes
    $L_x=100,200, 400$ collapse onto each other. Also shown is the
    analytical formula (line) described in the text.}
  \label{fig:omegak}
\end{figure}

\begin{figure}
\vspace{-5mm}
  \includegraphics[scale=\myfigscale]{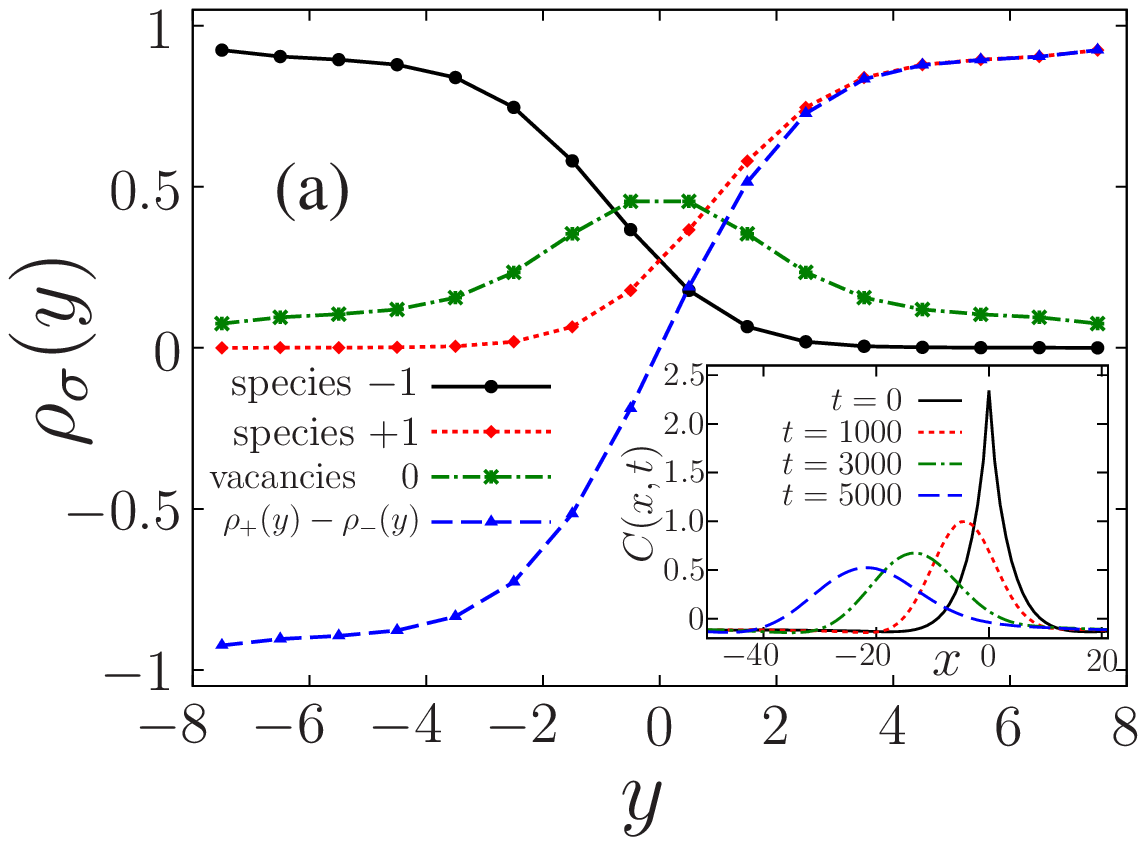}
  \includegraphics[scale=\myfigscale]{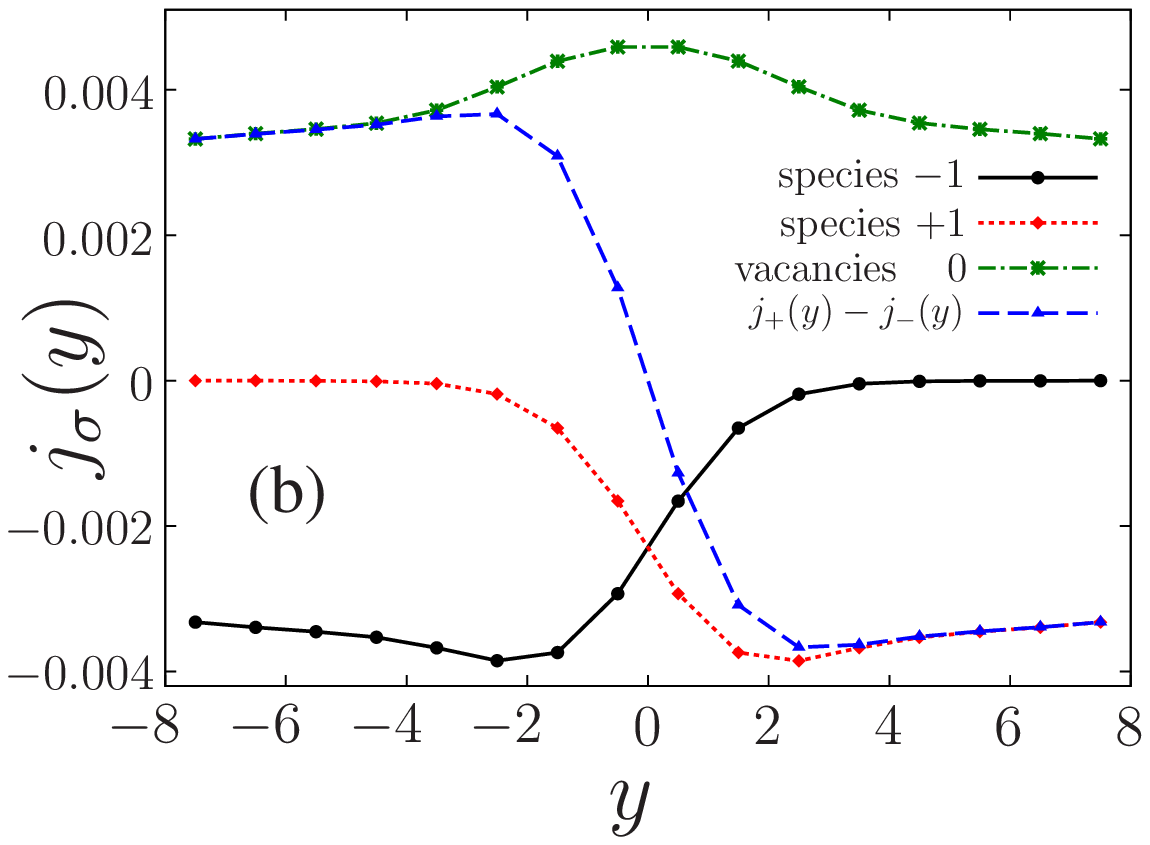}
  \caption{(color online) a) Density profiles $\rho_\sigma(y)$ of
    species $-1$, $+1$, and vacancies (0) for the Blume-Capel model
    under homogeneous drive of strength $f=0.25$ and $T=0.75 J/k_B$ as
    a function of the scaled distance from the mean interface position
    $y/L_y$. Also shown is the order parameter profile
    $\Phi(y)=\rho_+(y)-\rho_-(y)$.  Inset: $C(x,t)$ as a function of
    $x$ for various values of $t$ (as indicated).  b) Particle current
    profiles $j_\sigma(y)$ for species $\sigma=-1,+1,0$ and order
    parameter current profile, $\jop(y)=j_+(y)-j_-(y)$, as a function
    of $y$.}
  \label{fig:bc}
\end{figure}

In Fig.~\ref{fig:cxt}a we plot $C(x,t)$ as a function of $x$ for
several fixed values of time difference $t$ for the case of linear
drive, $\linconst=1$.  At $t=0$, a cusp is apparent at $x=0$ and
$C(x,0)$ exhibits long-ranged decay with distance $x$, characterizing
the (equal-time) spatial correlations of the interfacial fluctuations.
Upon increasing time, $t>0$, the position of the peak moves towards
negative values of $x$, its height decays and its width
increases. This behaviour clearly indicates the existence of damped
propagating modes that move in the negative $x$-direction. The
position of the maximum of $C(x,t)$ varies linearly with time. The
inferred velocity is $v_{\rm peak}=0.009$ (in units of lattice
constant per MC step); for not too large values of $\linconst$ the
velocity $v_{\rm peak}$ grows linearly with increasing $\linconst$.
For step-like drive of strength $f$, i.e., $F(y) \equiv f\, {\rm
  sgn}(y)$, where ${\rm sgn}(\cdot)$ is the sign function, we observe
very similar behaviour of $C(x,t)$ (data not shown), from which we
conclude that the occurrence of the interfacial motion is not tied to
the specific functional form of the (odd) driving field.
Qualitatively different behaviour occurs for even symmetry of the
drive, $F(y)=F(-y)$, such that the drive acts in the same direction
throughout the system. We find, for the cases considered, that the
order parameter current profile also attains even symmetry,
$\jop(y)=\jop(-y)$. For a V-shaped spatial dependence, $F(y) \equiv
\linconst|y|$, simulation data, shown in the inset of
Fig.~\ref{fig:cxt}a, show that with increasing time the peak of
$C(x,t)$ decreases in magnitude but remains stationary at $x=0$, which
indicates the absence of propagating modes. This observation holds
also for the case of uniform drive, $F(y)\equiv f={\rm const}$. The
intercept of the equal-time correlation function, $C(0,0) = \langle
h(0,0)^2 \rangle$, provides a measure of the (squared) interfacial
width \cite{SVAM}; we find for all cases considered that $C(0,0)$
decreases under drive. Comparison of the results for linear and
V-shaped drive indicates that the suppression of roughness \cite{zia}
is less strong in cases where interfacial motion occurs. The
interfacial transport is intimately related to a broken symmetry under
space reflection, $y \to -y$ and $x \to -x$, and species inversion,
$\sigma_i\to-\sigma_i$, that occurs for cases of odd driving.  As a
consequence, on average $-1$ spins in the region with $y>0$ move in
the {\em same} (negative) $x$ direction as $+1$ spins do in the region
with $y<0$. Hence ``intruders'' of the other phase move in the same
direction throughout the system. This direction is opposite to that of
the ``velocity profile'' of the order parameter,
$v_\phi(y)=\jop(y)/\Phi(y)$, which is $0\leq v_\Phi(y)\leq 0.022$ for
all $y$. See the inset of Fig.\ \ref{fig:cxt}b for an illustration of
the transport phenomenon.

The discrete Gaussian model \cite{chuiweeks} provides a reduced
description with interfacial degrees of freedom only. Nevertheless
these are known to exhibit long-wavelength equilibrium fluctuations
that are characteristic of real fluid interfaces. In a one-dimensional
system the interface is represented by integer height variables $h(x)$
with $x=1, \ldots, L_x$ and periodic boundary conditions. The
Hamiltonian is ${\cal H}_{\rm DG}= (J/2) \sum_{x=1}^{L_x}
\left(h(x+1)-h(x)\right)^2$.  In our (conserved) dynamics a site $x$
and one of its neighbours $x'=x\pm 1$ are chosen at random, and the
heights are changed as $h(x) \to h(x)+1$ and $h(x') \to h(x')-1$, with
probability given by the modified Metropolis rate. This can be viewed
as an (effective) biased diffusion step of a particle along the
interface from $x'$ to $x$. We have performed MC simulations for the
same parameters as for the Ising model, and for two cases of
driving. In the first case, the work is $\Delta W=-J\linconst (x'-x)
(h(x)+h(x'))/2$, corresponding to linear drive in the Ising lattice
gas. As shown in Fig.~\ref{fig:cxt}b, the temporal height-height
correlation functions exhibit characteristics very similar to those
found in the microscopic (Ising) model with capillary wave motion.
The velocity of the position of the maximum is $v_{\rm peak}=0.08$ (in
units of lattice constants per MC step, where one MC step consists of
$L_x$ exchange moves), significantly larger than that of the Ising
model.  Biasing the moves to the left (say) irrespective of the height
variable, $\Delta W= -J(x'-x) f$, corresponds to uniform drive in the
Ising model. As in the Ising model, no transport is observed. In cases
where transport occurs, the combined symmetry $h\to -h$, $x\to -x$,
and exchange of $x$ and $x'$, is broken.

The short-time dynamics of the capillary waves can be characterized by
a dispersion relation of frequency $\omega$ as a function wave number
$q=2\pi k/L_x$, obtained from the (average) phase shift of mode
$k=1,2,\ldots, L_x$ in unit time as $\omega(q)= \arg \langle \tilde
h^\ast(q,t) \tilde h(q,t+dt) \rangle/dt$, where $\tilde h(q,t)$ is the
spatial Fourier transform of $h(x,t)$, and $dt$ is a small time
interval ($L_x/10$ spin exchanges).  Surprisingly, for small
wavenumbers $q$ the driven Ising lattice possesses a nonlinear
dispersion relation, see Fig.~\ref{fig:omegak}a where results are
shown for the case of linear drive with $\linconst=1$. In contrast,
$\omega(q)$ for the discrete Gaussian model is linear for small $q$,
see Fig.~\ref{fig:omegak}b.  Modelling the dynamics by a simple linear
transport operator $\partial_t- v\partial_x$, with the continuous
partial time derivative $\partial_t$, discrete spatial derivative
$\partial_x h(x,t)=[h(x+1,t)-h(x-1,t)]/2$, and plane wave modes
$\exp({\rm i}(\omega t+qx))$ yields $\omega(q)=v \sin(q)$. This
describes the simulation data very well, see
Fig.\ \ref{fig:omegak}b. For small $q$, the behaviour $\omega(q)=vq$,
with $v=0.0760(1)$, is in good agreement with the value of $v_{\rm
  peak}$ obtained from analysis of $C(x,t)$.  The much richer
behaviour of the Ising model can be fitted to
$\omega(q)=(v+2u)\sin(q)-u\sin(2q)+s\sin^2(q)$ with small-$q$
expansion $\omega(q)=vq+sq^2+(u-v/6)q^3$, and $v=0.0172(7),
u=0.0402(2), s=0.0263(13)$. A linear transport operator
$\partial_t-v\partial_x+u\partial_x^3$ (5-point stencil for
$\partial_x^3$) yields the first and second terms of the fit function,
but the third term cannot be obtained from a linear transport equation
with real coefficients \cite{footnoteAmplitude}. We expect a
quantitative description of the full dynamics, as measured by
$C(x,t)$, to require a non-linear description.

In the kinetic Blume-Capel model the presence of vacancies (sites $i$
with $\sigma_i=0$) creates richer microscopic dynamics. In contrast to
the Ising model, where driving $+1$ particles is intimately related to
counter-driving $-1$ particles in the opposite direction, the
Blume-Capel model offers the possibility to {\em co-drive} the $\pm 1$
species in the same direction.  The corresponding driving fields
$F_\sigma(y)$ acting on species $\sigma=\pm 1$ then obey
$F_+(y)=F_-(y)\equiv F(y)$.  The work term for an exchange of spins
$i$ and $l$ with coordinates $x$ and $x'=x+1$, respectively, is
$\Delta W= -JF(y)(\sigma_i^2-\sigma_l^2)$. Simulation results for
vacancy concentration of 20\%, where we find the confined interface to
be stable, indicate that the symmetries of the driving field $F(y)$
and of $\jop(y)$ are no longer the same, e.g., that an even field can
give rise to an odd order parameter current profile, see
Fig.~\ref{fig:bc} for results of the density profile,
$\rho_\sigma(y)$, and current profile, $j_\sigma(y)$, of each species
$\sigma=-1,0,+1$ for the case of uniform drive, $F(y)={\rm
  const}$. Indeed motion of capillary waves is observed, see the inset
of Fig.\ \ref{fig:bc}a for results for $C(x,t)$. We do not find wave
motion when counter-driving, i.e.\ for $F_+(y)=-F_-(y)$. All these
findings are fully consistent with the proposed scenario that the
interfacial motion is caused by the symmetry of $\jop(y)$ and not by
that of the driving fields.

For general cases where the driving field and hence the order
parameter current profile do not possess unique but rather mixed
spatial symmetry, we find, for the Ising model, that it is the
presence of an odd component that leads to interfacial motion.  We
have also checked that the transport phenomenon is not specific to
two-dimensional systems, but also occurs in the three dimensional
Ising model, where simulation results indicate that the direction of
motion is parallel to that of the drive and that the second coordinate
parallel to the interface plays a mere spectator role.

In view of possible microfluidic applications, it would be very
interesting to test the validity of the proposed scenario in
experiments, e.g., using colloidal dispersions \cite{derks}, and in
molecular dynamics simulations \cite{tarazona}.

We thank D. Abraham, S. Dietrich, A. Gambassi, and T. Fischer for
useful discussions and the EPSRC for support under Grant
No.\ EP/E065619.

\end{document}